\documentclass[prd, aps, twocolumn, reprint, amsfonts, amssymb, amsmath, preprintnumbers, showpacs, nofootinbib, superscriptaddress]{revtex4-2}
\usepackage{graphicx, multirow,soul}
\usepackage[citecolor=blue]{hyperref}
\usepackage[all]{hypcap}
\usepackage{url}
\usepackage{color}
\usepackage{subfigure}
\usepackage{slashed}
\usepackage{float}
\usepackage{bbm}
\usepackage{bbold}
\usepackage{ulem}
\usepackage{lipsum}
\newcommand{\equaref}[1]{Eq.~(\ref{#1})}

\newcommand{\figref}[1]{Fig.~\ref{#1}}

\newcommand{\secref}[1]{Section~\ref{#1}}

\usepackage{tikz}
\tikzset{node distance=2cm, auto}
 \usepackage[compact]{titlesec}
 \titlespacing{\section}{0pt}{2ex}{1ex}
 \titlespacing{\subsection}{0pt}{1ex}{0ex}
  \titlespacing{\subsubsection}{0pt}{0.5ex}{0ex}

\allowdisplaybreaks

\begin{document}

\title{Gravitational waves and proton decay: complementary windows into GUTs}

\author{Stephen F. King}
\email[]{king@soton.ac.uk}
\email[]{https://orcid.org/0000-0002-4351-7507}
\affiliation{School of Physics and Astronomy, University of Southampton, Southampton, SO17 1BJ, U.K.}
\author{Silvia Pascoli}
\email[]{silvia.pascoli@durham.ac.uk}
\email[]{https://orcid.org/0000-0002-2958-456X}
\affiliation{Institute for Particle Physics Phenomenology, Department of
Physics, Durham University, South Road, Durham DH1 3LE, U.K.}
\author{Jessica  Turner}
\email[]{jessica.turner@durham.ac.uk}
\email[]{https://orcid.org/0000-0002-9679-5252}
\affiliation{Theoretical Physics Department, Fermi National Accelerator Laboratory, P.O. Box 500, Batavia, IL 60510, USA.}
\affiliation{Institute for Particle Physics Phenomenology, Department of
Physics, Durham University, South Road, Durham DH1 3LE, U.K.}
\author{Ye-Ling Zhou}
\email[]{ye-ling.zhou@soton.ac.uk}
\email[]{https://orcid.org/0000-0002-3664-9472}
\affiliation{School of Physics and Astronomy, University of Southampton, Southampton, SO17 1BJ, U.K.}

\date{\today}

\preprint{FERMILAB-PUB-20-187-T} 
\preprint{IPPP/20/20} 

\begin{abstract}
Proton decay is a smoking gun signature of  Grand Unified Theories (GUTs).  
Searches by Super-Kamiokande have resulted in stringent limits on the GUT symmetry breaking scale.
The large-scale multipurpose neutrino experiments DUNE, Hyper-Kamiokande and JUNO will 
either discover proton decay or further push the symmetry breaking scale above $10^{16}$ GeV.  Another possible observational consequence of GUTs is the formation of a cosmic string network produced
during the breaking of the GUT to the Standard Model gauge group. The evolution of 
such a string network in the expanding Universe produces a stochastic background of gravitational waves  which will be tested by a number of gravitational wave detectors 
over a wide frequency range. 
We demonstrate the non-trivial complementarity between the observation of
proton decay and gravitational waves produced from cosmic strings in determining  $SO(10)$ GUT breaking chains.
We show that such observations could exclude $SO(10)$
breaking via flipped $SU(5)\times U(1)$ or standard $SU(5)$,
while breaking via a Pati-Salam intermediate symmetry,
or standard $SU(5)\times U(1)$, may
be favoured if a large separation of energy scales associated with
proton decay and cosmic strings is indicated.
We note that recent results by the NANOGrav experiment have been interpreted as evidence for cosmic strings at a scale $\sim 10^{14}$~GeV. This would strongly point towards the existence of GUTs, with $SO(10)$ being the prime candidate. We show that the combination with already available constraints from proton decay allows to identify preferred symmetry breaking routes to the Standard Model.

\end{abstract}
\maketitle
%%%%%%%%%%%%%%%%%%%%%%%%%%%%%%%%%%%%%%%%%%%%%%%%%%%%%%%%%%%%%%%%%%%%%%%%%%%%%%%
%%%%%%                          Introduction                             %%%%%%
%%%%%%%%%%%%%%%%%%%%%%%%%%%%%%%%%%%%%%%%%%%%%%%%%%%%%%%%%%%
\section{INTRODUCTION}

Grand Unified Theories (GUTs) combine the strong, weak and electromagnetic forces of the Standard Model (SM) into a simple gauge group under which the fermions transform.
In such a framework, a larger underlying gauge symmetry is broken to the SM gauge group, $G_{\rm SM}=SU(3)_C \times SU(2)_L \times U(1)_Y$, either directly or via some symmetry breaking pattern. Following the Pati-Salam~\cite{Pati:1973uk} and $SU(5)$ \cite{Georgi:1972cj} proposals, many models have been  considered. Of particular interest are the $SO(10)$ GUTs~\cite{Fritzsch:1974nn} which predict neutrino masses and mixing and are based on a simple gauge group. 

A well known phenomenological prediction of GUTs 
is proton decay 
\cite{Weinberg:1979sa,Wilczek:1979hc,Weinberg:1980bf,Weinberg:1981wj,Sakai:1981pk,Dimopoulos:1981dw,Ellis:1981tv}.
 Super-Kamiokande has set stringent constraints on typical decay channels such as  $p\to \pi^0 e^+$ and $K^+ \bar{\nu}$ with the proton lifetime exceeding $10^{34}$ years~\cite{Abe:2014mwa,Miura:2016krn}.
 There are even more exciting prospects  
during the current decade thanks 
to the upcoming large-scale neutrino experiments, namely DUNE \cite{Acciarri:2015uup}, Hyper-Kamiokande \cite{Abe:2018uyc} and JUNO  \cite{An:2015jdp}.  

Another generic consequence of GUTs is the production of topological defects  when the GUT undergoes spontaneous symmetry breaking (SSB) 
 \cite{Jeannerot:2003qv}. 
Some of these, such as monopoles, need to be inflated away in order not to overclose the Universe. However, cosmic strings associated with the breaking of a $U(1)$ symmetry, which can be a gauged subgroup of the GUT \cite{Vilenkin:1984ib}, can remain until late times and have observational consequences. 
These cosmic strings (cs) 
are expected to produce 
gravitational waves (GWs) via the scaling of the string network \cite{Vilenkin:1984ib,Caldwell:1991jj,Hindmarsh:1994re}. 
These signals form a stochastic GW background (SGWB) today with an abundance proportional to the square of the $U(1)$ SSB scale, $\Lambda_{\rm cs}$. 
The observation of such events provides a unique probe of physics  at remarkably high scales
and has been recently considered in the context of 
 leptogenesis  \cite{Dror:2019syi} 
and GUTs~\cite{Buchmuller:2019gfy}. 

In this {\it Letter} we discuss the non-trivial complementarity between observing proton decay 
and GWs produced from cosmic strings in GUTs. In particular, we focus on the implications for determining possible $SO(10)$ GUT breaking chains. While searches for proton decay (pd) set a lower bound on the associate scale $\Lambda_{\rm pd}$ of new physics, 
the GW observations will place an upper bound on $\Lambda_{\rm cs}$. Moreover, 
we assume an inflationary epoch, at scale $\Lambda_{\rm inf}$, to eliminate unwanted topological defects. 
We explore the role of experimental searches 
in determining these three scales: $\Lambda_{\rm cs}$, $\Lambda_{\rm pd}$ and $\Lambda_{\rm inf}$.

In Section~\ref{sec:review}, we compare the scale of proton decay
and cosmic string formation
for breaking chains of $SO(10)$.
The synergy between observation of proton decay and GWs is 
discussed quantitatively in all possible $SO(10)$ breaking chains
in Section~\ref{sec:complementarity}.
We summarise and discuss our results in Section~\ref{sec:conclusion}.

%%%%%%%%%%%%%%%%%%%%%%%%%%%%%%%%%%%%%%%%%%%%%%%%%%%%%%%%%%%
%%%%%%                        GUT SIGNATURES												 %%%%%%
%%%%%%%%%%%%%%%%%%%%%%%%%%%%%%%%%%%%%%%%%%%%%%%%%%%%%%%%%%%
\section{TERRESTRIAL AND COSMIC SIGNATURES OF GUTS}  \label{sec:review}
\begin{figure}[t]
\begin{center}
        \includegraphics[width=.45\textwidth]{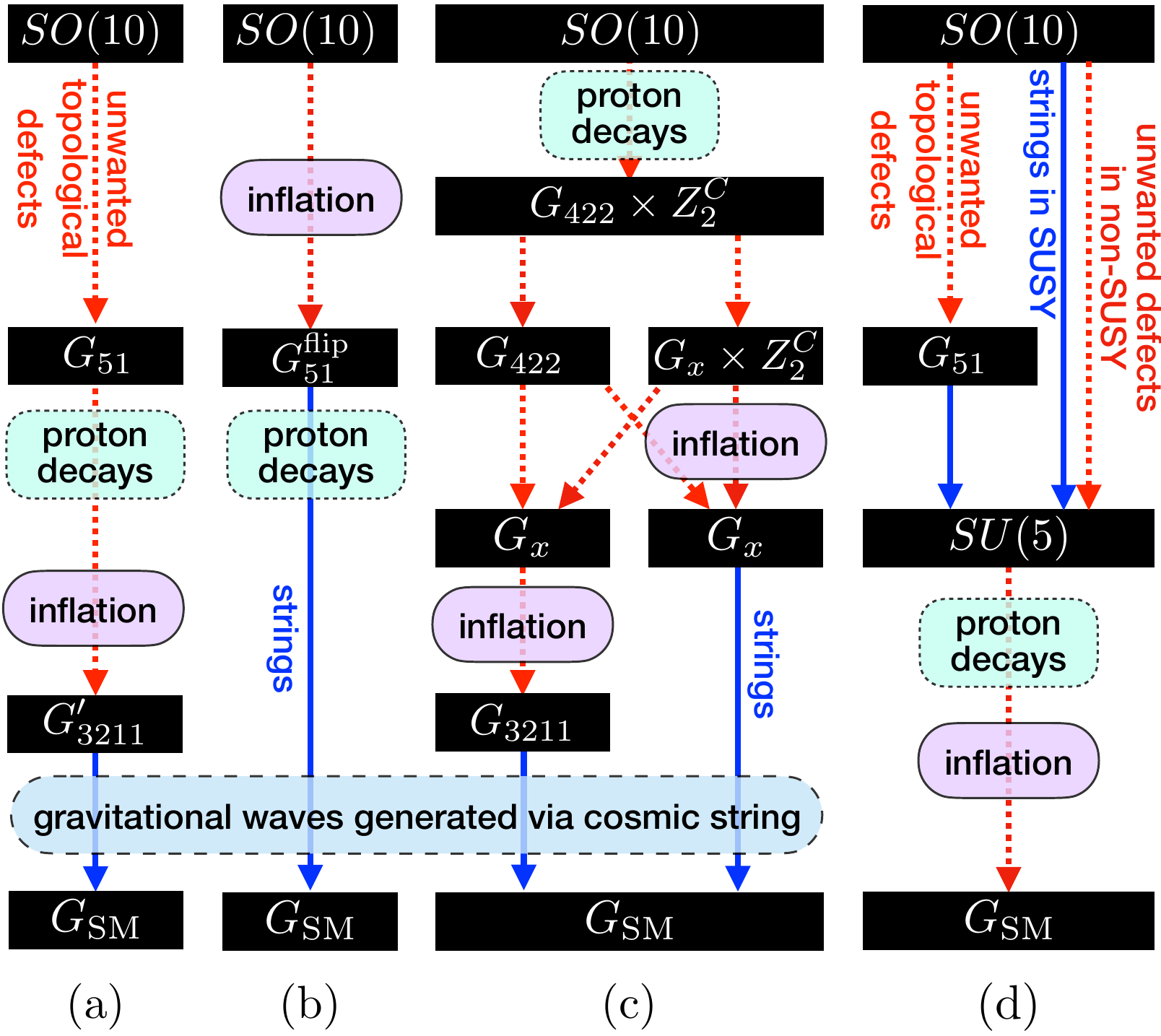}
        \label{fig:chains}
\caption{The breaking chains of $SO(10)$ to $G_{\rm SM}$ are shown along with their terrestrial and cosmological signatures
 where $G_x$ represents either $G_{3221}$ or $G_{421}$.
Defects with only cosmic strings (including cosmic string generated from preserved discrete symmetries) are denoted as blue solid arrows. Those including unwanted topological defects (monopoles or domain walls) are indicated by red dotted arrows.
The instability of embedded strings is not considered.
Removing an intermediate symmetry may change the type of unwanted topological defect but will not eliminate them. 
The highest possible scale of inflation, which removes 
unwanted defects, is assumed in this diagram.} \label{fig:breaking_1}
\end{center} 
\vspace{-2.5em}
\end{figure}

$SO(10)$ is the minimal simple GUT which offers the possibility of cosmic string generation. 
Its breaking to the SM gauge group can proceed along one of the breaking chains shown in Fig.~\ref{fig:breaking_1}, with the additional option of removing intermediate steps. We use the following abbreviations for the symmetries at an intermediate scale: 
\begin{eqnarray}\label{eq:symmetries}
&&G_{51} = SU(5) \times U(1)_X \,, \quad
G_{51}^{\rm flip} = SU(5)_{\rm flip} \times U(1)_{\rm flip} \,, \nonumber\\
&&G_{3221} = SU(3)_C \times SU(2)_L \times SU(2)_R \times U(1)_{B-L} \,, \nonumber\\
&&G_{3211} = SU(3)_C \times SU(2)_L \times U(1)_R \times U(1)_{B-L} \,, \nonumber\\
&&G_{3211}' = SU(3)_C \times SU(2)_L \times U(1)_Y \times U(1)_{X} \,, \nonumber\\
&&G_{421} = SU(4)_C \times SU(2)_L \times U(1)_Y\,, \nonumber\\
&&G_{422} = SU(4)_C \times SU(2)_L \times SU(2)_R \,.
\end{eqnarray}
Note that $G_{3211}$ and $G_{3211}'$ are  equivalent \cite{King:2017cwv}.
All possible  $SO(10)$ cases can be classified  into four types denoted as (a), (b), (c) and (d) in Fig.~\ref{fig:chains}. Types (a), (b) and (c) are models broken via standard $SU(5)\times U(1)$, flipped $SU(5)\times U(1)$\cite{Barr:1981qv,Derendinger:1983aj,DeRujula:1980qc,Antoniadis:1989zy}
and Pati-Salam $G_{422}$ \cite{Pati:1974yy}
respectively. Cases with standard $SU(5)$ \cite{Georgi:1972cj} as the lowest intermediate symmetry, are classified as type (d).
The scales of proton decay $\Lambda_{\rm pd}$ and 
cosmic strings $\Lambda_{\rm cs}$ are important testable parameters %which we  discuss 
discussed in the following.

%%%%%%%%%%%%%%%%%%%%%%%%%%%%%%%%%%%%%%%%%%%%%%%%%%%%%%%%%%%
%%%%%%                        PROTON DECAY												 %%%%%%
%%%%%%%%%%%%%%%%%%%%%%%%%%%%%%%%%%%%%%%%%%%%%%%%%%%%%%%%%%%
%%%%%%%%%%%%%%%%%%%%%%%%%%%%%%%%%%%%%%%%%%%%%%%%%%%%%%%%%%%
{\bf A. Proton Decay in \boldmath{$SO(10)$}}. 
As quarks and leptons are arranged in common multiplets in GUTs, heavy new states which  mediate baryon-number-violating (BNV) interactions are introduced. At low energy scales, these heavy states are integrated out and this induces 
 higher-dimensional BNV operators which lead to proton decay.

In the main body of the text, we will focus on non-supersymmetric contributions, while discussions on additional sources provided by supersymmetric extensions will be discussed in the Supplemental Material. In summary, SUSY with R-parity has similar phenomenological/cosmological  consequences, see Fig.~\ref{fig:breaking_1}, with the addition of the $K^+ \bar{\nu}$ proton decay channel. 

At low energy, the most important operators which respect $G_{\rm SM}$ are the dimension-six ones arising from gauge contributions,
\begin{eqnarray} \label{eq:D6}
&&\frac{\epsilon_{\alpha\beta}}{\Lambda_1^2} \left[
(\overline{u_R^{c}} \gamma^\mu Q_\alpha)(\overline{d_R^{c}} \gamma_\mu L_\beta) +
(\overline{u_R^{c}} \gamma^\mu Q_\alpha)(\overline{e_R^{c}} \gamma_\mu Q_\beta) \right ]
\\
&+&
\frac{\epsilon_{\alpha\beta}}{\Lambda_2^2} \left[
(\overline{d_R^{c}} \gamma^\mu Q_\alpha)(\overline{u_R^{c}} \gamma_\mu L_\beta) +
(\overline{d_R^{c}} \gamma^\mu Q_\alpha)(\overline{\nu_R^{c}} \gamma_\mu Q_\beta) \right ], \nonumber
\end{eqnarray}
where 
$\alpha, \beta$ denote $SU(2)_L$ indices and
$\Lambda_1$, $\Lambda_2$ are the UV-complete scales of the GUT symmetry  \cite{Weinberg:1979sa,Wilczek:1979hc,Weinberg:1980bf,Weinberg:1981wj,Sakai:1981pk}. 
 For types (a) and (d), $\Lambda_1$ and $\Lambda_2$ correspond to the $SU(5)$ and $SO(10)$ breaking 
 scales, respectively, and thus $\Lambda_1 < \Lambda_2$.  While for type (b), $\Lambda_2< \Lambda_1$ and 
 $\Lambda_1 = \Lambda_2$ for type (c). 
In general, the lower of these two scales will mediate the dominant proton decay channel and we indicate it as $\Lambda_{\rm pd}$.

These operators induce a series of proton decay channels. The most stringently constrained is $p \to \pi^0 e^+$ as determined by Super-Kamiokande, $\tau_{\pi^0 e^+} > 1.6 \times 10^{34}$ years ($90\%$ C.L.,  100\% branching ratio assumed) \cite{Miura:2016krn}.
This bound translates to the lower limits of  $\Lambda_1 >  6.7 \times 10^{15}$~GeV and $\Lambda_2 >  3.9 \times 10^{15}$~GeV, respectively, using $\tau_{\pi^0 e^+} \simeq 8\times 10^{34}\,{\rm years} \times (\Lambda_{\rm 1}/10^{16}\,{\rm GeV})^4$ \cite{Murayama:2001ur} or $7\times 10^{35}\,{\rm years} \times (\Lambda_{\rm 2}/10^{16}\,{\rm GeV})^4$ \cite{Ellis:2020qad}, respectively. 
Hyper-Kamiokande offers at least an order of magnitude improvement  
\cite{Abe:2018uyc} which will further push the lower bound of $\Lambda_1$ above $10^{16}$~GeV.

%%%%%%%%%%%%%%%%%%%%%%%%%%%%%%%%%%%%%%%%%%%%%%%%%%%%%%%%%%%
%%%%%%%%%%%%%%%%%%%%%%%%%%%%%%%%%%%%%%%%%%%%%%%%%%%%%%%%%%%

%%%%%%%%%%%%%%%%%%%%%%%%%%%%%%%%%%%%%%%%%%%%%%%%%%%%%%%%%%%
%%%%%%                        GUTS COSMIC STRINGS AND GWS								%%%%%%
%%%%%%%%%%%%%%%%%%%%%%%%%%%%%%%%%%%%%%%%%%%%%%%%%%%%%%%%%%%
{\bf B. Gravitational Waves From Cosmic Strings}. 
The cosmological consequence of SSB from the GUT to the SM gauge group is the formation of topological defects. These defects generically  arise from 
the breaking of a group, $G$, to its subgroup, $H$, such that a manifold of equivalent vacua, $M \simeq G/H$, exists. 
Monopoles form when the manifold $M$ contains non-contractible two-dimensional spheres, cosmic strings when it contains non-contractible loops 
and domain walls when $M$ is disconnected.
Different GUT breaking chains result in different combinations of topological defects forming at various scales; these
have been comprehensively categorised in  \cite{Jeannerot:2003qv}  where it was shown that the vast majority of GUT breaking chains produce cosmic strings.
In \figref{fig:chains}, we summarise all  possible symmetry breaking chains and associated defects as derived in  Ref.~\cite{Jeannerot:2003qv}.
We note that embedded strings can be generated if a $Z_2$ symmetry is preserved \cite{Kibble:1982ae}; however, we do not distinguish them from topological strings and both scenarios are indicated by the 
blue lines of \figref{fig:chains}.

Cosmic strings are a source of GWs as they actively perturb the metric at all times.
If cosmic strings form after inflation, they exhibit a scaling behaviour where the stochastic GW
spectrum is relatively flat as a function of the frequency and the amplitude is proportional to
the string tension $\mu$. We refer to the string formation scale as $\sqrt{\mu}\equiv \Lambda_{\rm cs}$  as, without fine-tuning, all gauge coefficients in GUTs are of order one. We note that this scale is identical to the symmetry breaking scale up to an order one coefficient. This scale, if exists, is the lowest intermediate scale of $SO(10)$ GUT breaking, as indicated in Fig.~\ref{fig:breaking_1}.
The GWs are sourced when the cosmic strings intersect to form loops. 
Cusps on these strings emit strong beams of 
high-frequency GWs or {\it bursts}, that constitute a SGWB if unresolved over time \cite{Damour:2001bk,Damour:2004kw}.
An inflationary period can 
suppress the SGWB in high frequencies \cite{Guedes:2018afo}. 
However, it was recently shown that cosmic string network regrowth can occur to the extent that its associated GW signal is observable \cite{Cui:2019kkd}, contrary to what was naively expected. This string regrowth is contingent upon the initial number of cosmic strings per Hubble volume and the number of e-folds into inflation that  the string formation occurs. A detailed discussion of these initial conditions and up-to-date sensitivities of the GW observatories on the string tension are provided in the aforementioned reference.

%%%%%%%%%%%%%%%%%%%%%%%%%%%%%%%%
\begin{figure}[t]
\centering
        \includegraphics[width=.45\textwidth]{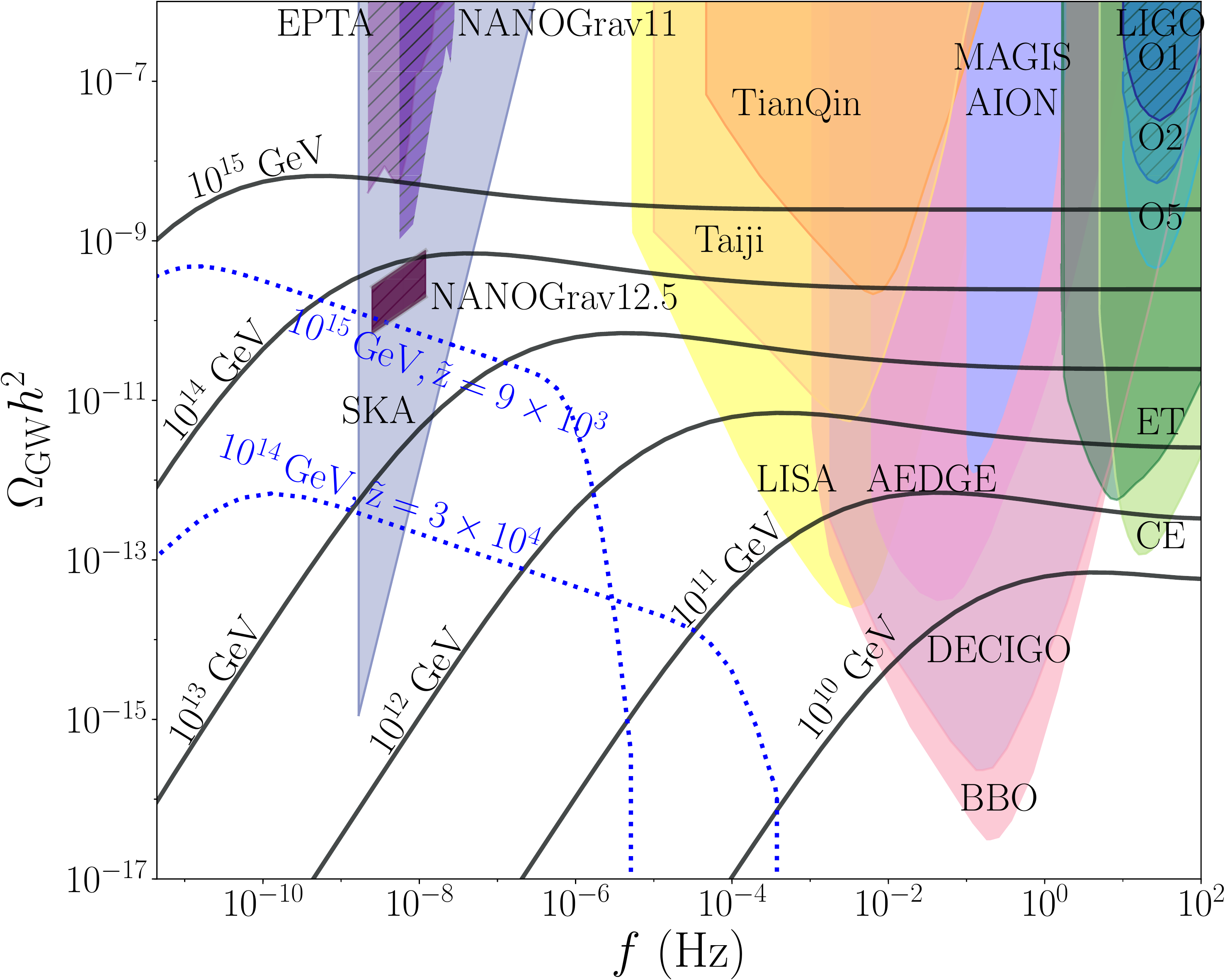}
        \label{fig:gull}
\caption{
SGWB predicted from undiluted (solid black) and diluted (dashed bule) cosmic string networks, where  $\Lambda_{\rm cs} = 10^{10,11,\cdots,15}$~GeV are input. $\tilde{z}$ denotes  the redshift when strings return to the horizon, namely $H(\tilde{z})L(\tilde{z}) = 1$. Current  (hatched) and future (coloured) experimental limits are shown as comparision.} 
\label{fig:GWs}
\vspace{-1.5em}
\end{figure}

In \figref{fig:GWs} we show sensitivities of current and future GW experiments alongside  the predicted SGWB for cosmic strings undiluted (solid curves) and diluted (dashed curves) by inflation.
The $U(1)$ symmetry breaking scale  $\Lambda_{\rm cs} = 10^{10,11,\cdots,15}$~GeV corresponds to  $G \mu\simeq 0.7\times 10^{-18,-16,\cdots,-8}$, respectively, where $G$ is Newton's constant. 
We provide formulations of SGWB in both the undiluted and diluted cosmic strings scenarios in the Supplementary Material, following \cite{Blanco-Pillado:2017oxo,Cui:2018rwi} and \cite{Cui:2019kkd}, respectively. Furthermore, for a comprehensive review on cosmic strings see Ref. \cite{Vilenkin:2000jqa} and references therein. 

Applying these standard assumptions, a large range of $\Lambda_{\rm cs}$ can be explored using GW detectors. LIGO O2 \cite{LIGOScientific:2019vic} has excluded cosmic strings formation at $\Lambda_{\rm cs} \sim 10^{15}$~GeV in the high frequency regime $10$-$100$~Hz.  While in low frequency band, $1$-$10$~nHz, the null result of EPTA \cite{Lentati:2015qwp} and NANOGrav 11-year data \cite{Arzoumanian:2018saf} constrains the upper bound of $\Lambda_{\rm cs}$ below $10^{15}$~GeV and $10^{14}$~GeV, respectively.\footnote{These constraints could be relaxed due to the choice of prior as recently pointed in \cite{Hazboun:2020kzd}.}
Planned pulsar timing arrays SKA \cite{Janssen:2014dka}, space-based laser interferometers LISA \cite{Audley:2017drz}, Taiji \cite{Guo:2018npi}, TianQin \cite{Luo:2015ght}, BBO \cite{Corbin:2005ny}, DECIGO \cite{Seto:2001qf}, ground-based interferometers Einstein Telescope \cite{Sathyaprakash:2012jk} (ET), Cosmic Explorer \cite{Evans:2016mbw} (CE), and atomic interferometers MAGIS \cite{Graham:2017pmn}, AEDGE \cite{Bertoldi:2019tck}, AION \cite{Badurina:2019hst} will probe $\Lambda_{\rm cs}$ values in a wide regime $10^{10\text{-}14}$~GeV.
As the spectrum of GWs produced via diluted cosmic strings decreases rapidly for $f > 10^{-6}$~Hz, this allows them  to be distinguished from the undiluted cosmic strings as shown in Fig.~\ref{fig:GWs}. 

Unwanted topological defects are generated in all $SO(10)$ breaking chains, as indicated in Fig.~\ref{fig:breaking_1} and inflation is a promising means to remove them. 
Consistent hybrid inflation models have been achieved via GUT breaking \cite{BasteroGil:2006cm,Pallis:2013dxa}.
The shape and magnitude of the inflaton potential are imprinted in the primordial density perturbations which
are characterised by the spectral index and the tensor-to-scalar ratio
in cosmic microwave background (CMB) measurements, from which
the upper limit on inflation is $\Lambda_{\rm inf}< 1.6 \times 10^{16}$ GeV (95\% C.L., Planck) \cite{Akrami:2018odb}. 
Future CMB measurements can improve the tensor-to-scalar ratio upper limit to $0.001$ (95\% C.L., CMB-S4) \cite{Abazajian:2019eic}, corresponding to $\Lambda_{\rm inf}<5.7\times 10^{15}$~GeV.
%%%%%%%%%%%%%%%%%%%%%%%%%%%%%%%%%%%%%%%%%%%%%%%%%%%%%%%%%%%
%%%%%%                        MAIN RESULTS								                                                         %%%%%%
%%%%%%%%%%%%%%%%%%%%%%%%%%%%%%%%%%%%%%%%%%%%%%%%%%%%%%%%%%%
\section{SYNERGY BETWEEN PROTON DECAY AND GW MEASUREMENTS}\label{sec:complementarity}
Planned future proton decay searches will either put a more stringent lower bound on $\Lambda_{\rm pd}$  or, in the presence of a signal, will provide further insight into the GUT symmetry structure. 
Due to the relatively model-independent  nature of the   operators shown in  \equaref{eq:D6}, the following experimental results
are of particular interest:
 \begin{itemize}
\item Proton decay is observed in the $\pi^0e^+$ channel. This provides an  explicit link between $\Lambda_{\rm pd}$ and $\tau_{\pi^0e^+}$. 
\item Proton decay is observed in the $K^+\bar{\nu}$ channel. This case provides a  weaker connection to $\Lambda_{\rm pd}$ due to the involvement of the unknown SUSY-breaking scale.
\end{itemize}

The observation of GWs from cosmic strings is crucially dependent on the scale of inflation. We consider two possibilities: i) the case of string formation after inflation, namely $\Lambda_{\rm cs}< \Lambda_{\rm inf}$, for which a SGWB is generated from undiluted strings; and ii) the case of GWs from diluted cosmic strings, if  $\Lambda_{\rm cs} \sim \Lambda_{\rm inf}$. The case $\Lambda_{\rm cs}> \Lambda_{\rm inf}$ will not be considered as there are no associated cosmological signatures of GUTs. 

From the synergy of experimental data discussed in \secref{sec:review} ($\Lambda_{\rm pd}\gtrsim 10^{15}$ GeV, $\Lambda_{\rm inf}<10^{16}$ GeV and $\Lambda_{\rm cs}<10^{14}$ GeV) certain ordering of scales are already excluded such as  $\Lambda_{\rm inf}>\Lambda_{\rm cs}\sim \Lambda_{\rm pd}$ and $\Lambda_{\rm inf} \gtrsim \Lambda_{\rm cs}> \Lambda_{\rm pd}$.\footnote{The latter is not predicted in $SO(10)$ but in enlarged symmetries such as $E_6$ \cite{Jeannerot:2003qv}.}
We first discuss the various scales for the
type (a) chain 
and then examine the remaining breaking chains. 
%%%%%%%%%%%%%%%%%%%%%%%%%%%%%%%%
%%%%%%      TYPE (A)                     %%%%%%%%%%%%%
%%%%%%%%%%%%%%%%%%%%%%%%%%%%%%%%

{\bf Type (a)} 
is characterised by $\Lambda_{\rm pd} > \Lambda_{\rm cs}$. The main source of proton decay is provided by $\Lambda_1$-suppressed operators in \equaref{eq:D6} \cite{Weinberg:1979sa,Wilczek:1979hc}.
A cosmic string network is produced at $\Lambda_{\rm cs}$. 
However, the observational signal of associated  GWs depends on $\Lambda_{\rm inf}$ as follows. 

As discussed, inflation must be introduced to remove unwanted defects produced in the first and second steps of the breaking. To achieve this, the inflationary scale $\Lambda_{\rm inf}$ should not be higher that the second-step breaking scale, $\Lambda_{\rm pd}$. 
Therefore, there  are three possible orderings of the relevant scales. 
1) $\Lambda_{\rm pd}  \gtrsim \Lambda_{\rm inf} >\Lambda_{\rm cs} $, proton decay may be observed  in conjunction with an undiluted GW signal, which is an ideal possibility from the experimental perspective. 2) $\Lambda_{\rm pd}  > \Lambda_{\rm inf} \sim \Lambda_{\rm cs}$, proton decay may be observed in combination with a diluted GW signal. 3) $\Lambda_{\rm pd}  > \Lambda_{\rm cs} > \Lambda_{\rm inf}$, proton decay could be observed but no associated GW signal is detected. 
%%%%%%%%%%%%%%%%%%%%%%%%%%%%%%%%%%
\begin{figure*}[t]
\begin{center}
        \includegraphics[width=.95\textwidth]{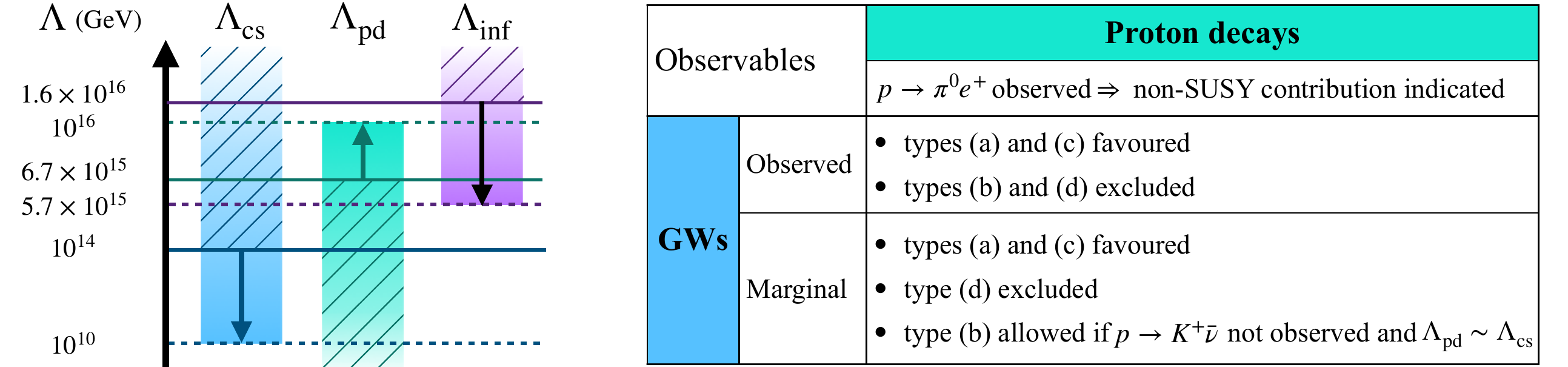}
        \label{fig:observations}
\caption{GUTs constrained by observations of GWs and proton decays. Left panel: Current (hatched) and future (solid) exclusion limits of energy scales of cosmic string formation, proton decays and inflation. 
$\Lambda_{\rm pd} \sim \Lambda_1$ is approximated 
and the exclusion limit of $\Lambda_{\rm cs}$ is shown in the undiluted case only. Right panel: Potential conclusions of GUT properties based on observations of GWs and proton decays in next-generation experiments.}
\end{center}
\vspace{-1.5em}
\end{figure*}

%%%%%%%%%%%%%%%%%%%%%%%%%%%%%%%%%%%%%%%%%%%%%%%%%%%%%%%%%%%
%%%%%%                   ALL REST SO(10) CHAINS  					                                				  %%%%%%
%%%%%%%%%%%%%%%%%%%%%%%%%%%%%%%%%%%%%%%%%%%%%%%%%%%%%%%%%%%
\textbf{Type (b)} is associated with flipped $SU(5)\times U(1)$ and proton decay proceeds dominantly via the pion channel.
Similarly to (a), 
string formation occurs in the final breaking step. 
This case is characterised by $\Lambda_{\rm pd} \sim \Lambda_{\rm cs}$.
Given the current limits on proton decay and GWs 
(which implies $\Lambda_{\rm pd}\gtrsim 10^{15}$ GeV and 
$\Lambda_{\rm cs}\lesssim 10^{14}$ GeV for  the undiluted cosmic string scenario) 
it may appear $\Lambda_{\rm pd} \sim \Lambda_{\rm cs}$ is already excluded. However, as before,
the observability of GWs depends on the scale of inflation
$\Lambda_{\rm inf}$ as we now discuss.

If the scale of inflation is high, then indeed
the scale ordering $\Lambda_{\rm inf}>\Lambda_{\rm pd}\sim\Lambda_{\rm cs}$ can already be excluded. 
However, 
$\Lambda_{\rm inf}\sim \Lambda_{\rm pd}\sim\Lambda_{\rm cs}$ remains viable as the 
SGWB produced from diluted strings is suppressed relative to the undiluted case. Given the sensitivities, this ordering can be tested in the next generation experiments.

\textbf{Type (c)} 
represents a class of cases which have the common feature that 
proton decay is associated with 
the breaking of $SO(10)$ to the Pati-Salam gauge group where cosmic strings are generated by the last step of  breaking.  
Hence $\Lambda_{\rm pd} >  \Lambda_{\rm cs}$ as in type (a).
As before, the observability of GWs depends on the scale of inflation
$\Lambda_{\rm inf}$, to which we turn.
The breaking of $G_{422}$  results in the production of unwanted defects at each stage of SSB prior to the final breaking that produces the string network. Therefore, $\Lambda_{\rm inf}$ must occur below the
breaking of $G_{422}$. Notwithstanding, the scale ordering of this class of models can be determined in a similar way to type (a).

To distinguish between type (a) and (c)  further specification of the model 
is required.
From this, predictions of nucleon decay branching ratios could be used to differentiate between the breaking chains (see {\it e.g.} Ref.~\cite{Nath:2006ut}). 
Furthermore, $\Lambda_{\rm pd}$ in type (c) chains can be significantly  higher than
$10^{16}$~GeV if there are threshold corrections from intermediate symmetries at low scale, {\it e.g.}, $10^{10 \text{-} 12}$~GeV \cite{Bertolini:2009qj,Chakrabortty:2019fov}. 
Such low scale SSB may be linked to the origin of neutrino masses and leptogenesis \cite{Pascoli:2016gkf,Long:2017rdo}. An observation of low scale GWs may favour some specific breaking chains of this type.

\textbf{Type (d)} has the same $SU(5)$ intermediate symmetry as type (a) and therefore similar predictions for proton decay as in type (a) but with $\Lambda_{\rm cs} > \Lambda_{\rm pd}$.
However, the inflation scale must be lower than the proton decay scale
$\Lambda_{\rm pd}  > \Lambda_{\rm inf}$,
since monopoles generated in the final step of symmetry breaking must be inflated away. Unfortunately, this also inflates away the cosmic strings. 
Hence, any associated GW detection via cosmic strings (diluted or undiluted) would exclude this class of breaking chains
under our assumption the GW signal is associated to the $SO(10)$ breaking.

Our analysis is summarised in Fig.~\ref{fig:observations}. In the right panel we tabulate 
how observing proton decay via the pion channel in conjunction with  GWs can be used to exclude or favour certain 
breaking chains and also provide information on the scale ordering.
The consequences of null observations  are not given  in Fig.~\ref{fig:observations}. In the event proton decay is not observed in the upcoming neutrino experiments, 
the limit on the UV-complete scale $\Lambda_{\rm pd}$ will be pushed even higher.  
On the other hand, future non-observation of cosmic string-induced GWs would suggest an inflationary era occurred after cosmic string formation. 
In addition, improved CMB measurements will allows  more stringent upper bound for 
$\Lambda_{\rm inf}$ to be placed which will in turn be an upper bound for $\Lambda_{\rm cs}$ if cosmic strings are to be observed. 
This is schematically shown in the left panel of Fig.~\ref{fig:observations} where coloured and 
hatched regions indicate current and future experimental limits to probe these scales. 
For example, future experiments may constrain
$\Lambda_{\rm pd} > \Lambda_{\rm inf}$. In SUSY $SO(10)$, the same scale orderings between $\Lambda_{\rm pd}$ and $ \Lambda_{\rm inf}$ can be obtained, although a less precise value of $\Lambda_\mathrm{pd}$ can be inferred from an observation $p \to K^+ \bar{\nu}$, see the Supplemental material.

Very recently, NANOGrav 12.5-year data finds strong evidence of SGWB with a power law spectrum in the frequency band $2.5$-$12$~nHz \cite{Arzoumanian:2020vkk}, as shown in Fig.~\ref{fig:gull}. It has been explained in the framework of string network scaling with $G\mu \sim (2 \times 10^{-11}, 2 \times 10^{-10})$ at 95\% CL \cite{Ellis:2020ena}, corresponding to $\Lambda_{\rm cs} \sim (0.5, 1.7) \times 10^{14}$~GeV.\footnote{Variations of string models such as small loops \cite{Blasi:2020mfx} and metastable strings \cite{Buchmuller:2020lbh} would point to an even higher string formation scale.} As we explained above, if confirmed, the combination with already available constraints from proton decay excludes the type (b) and type (d) breaking chains.
Moreover, it does not support a large class of type (c) ones. As indicated in \cite{Bertolini:2009qj,Chakrabortty:2019fov}, type (c) with one or two intermediate scales predict the lowest intermediate scale either below or marginally consistent with the NANOGrav lower bound $5 \times 10^{13}$~GeV. Therefore, a preference for Type (a) emerges and future information from proton decay experiments would crucially allow to further strenghten this conclusion.

%%%%%%%%%%%%%%%%%%%%%%%%%%%%%%%%%%%%%%%%%%%%%%%%%%%%%%%%%%%
%%%%%%                   SUMMARY 					                                						  %%%%%%
%%%%%%%%%%%%%%%%%%%%%%%%%%%%%%%%%%%%%%%%%%%%%%%%%%%%%%%%%%%
\section{SUMMARY AND CONCLUSION \label{sec:conclusion}}
We propose a strategy to use both proton decay and gravitational waves (GWs) as a means of identifying possible breaking chains of Grand Unified Theories (GUTs). We focus on $SO(10)$ GUT models
and categorise them according to their symmetry breaking patterns as shown in Fig.~\ref{fig:chains}(a)-(d), corresponding to 
standard $SU(5)\times U(1)$, flipped $SU(5)\times U(1)$, Pati-Salam
and standard $SU(5)$, respectively.

For each pattern of breaking, we compare the scale of proton decay, $\Lambda_{\rm pd}$, with the cosmic string formation scale, $\Lambda_{\rm cs}$.
These scales can have important testable consequences as they are related
to the proton lifetime and the generation of GWs via cosmic strings. 
The determination of these scales, in particular their ordering, provides useful information in assessing the viability of a given class of breaking chains within $SO(10)$ GUTs. 

Our results are summarised in Fig.~\ref{fig:observations}.
In particular, such observations could exclude $SO(10)$
breaking via flipped $SU(5)\times U(1)$ or standard $SU(5)$,
while breaking via a Pati-Salam intermediate symmetry,
or standard $SU(5)\times U(1)$, 
may be favoured if a large separation of energy scales associated with
proton decay and cosmic strings is indicated. 

We note that recent evidence of a stochastic background of gravitational waves by the NANOGrav experiment can be interpreted as due to cosmic strings at a scale $\sim 10^{14}$~GeV. This result would strongly point towards the existence of GUTs, with $SO(10)$ being the prime candidate. Our results show that the combination with already available information from proton decay can identify the symmetry breaking pattern down to the Standard Model, with strong preference for type (a) or a subset of type (c). 

In conclusion, we have entered an exciting era where new observations of GWs from the heavens and proton decay experiments from under the Earth can provide complementary windows 
to reveal the details of the unification of matter and forces at the highest energies.

%%%%%%%%%%%%%%%%%%%%%%%%%%%%%%%%%%%%%%%%%%%%%%%%%%%%%%%%%%%%%%%%%%%%%%%%%%%%%%%%%%%%%%%%%%%%%%%%%%%%%%%%%%%%%%%%%%%%%%%%%%%%
\acknowledgements

This work was partially supported by the STFC Consolidated Grant ST/L000296/1, the European Research Council under ERC Grant NuMass (FP7-IDEAS-ERC ERC-CG 617143), Fermi Research Alliance, LLC under Contract No. DE-AC02-07CH11359 with the U.S. Department of Energy and by the European Union’s Horizon 2020 research and innovation programme under the Marie Sklodowska-Curie grant agreement No. 690575 (RISE InvisiblesPlus) and No. 674896 (ITN Elusives). 
J.T would like to thank Nikita Blinov  and Holger Schulz for useful discussions.

\bibliographystyle{apsrev4-1}
\bibliography{ref}{}

\onecolumngrid
\newpage

\begin{center}
\textbf{\large Supplemental Material for \\
Gravitational waves and proton decay: complementary windows into GUTs }\\[.2cm]
Stephen F. King,$^1$ Silvia Pascoli,$^2$ Jessica  Turner,$^{3,2}$ Ye-Ling Zhou$^1$\\

{\it $^1$School of Physics and Astronomy, University of Southampton, Southampton, SO17 1BJ, U.K.

$^2$Institute for Particle Physics Phenomenology, Department of Physics, \\
Durham University, South Road, Durham DH1 3LE, U.K.

$^3$Theoretical Physics Department, Fermi National Accelerator Laboratory, P.O. Box 500, Batavia, IL 60510, USA.
}

\end{center}
%\section{NUMERICAL METHOD TO CALCULATE GWS FROM COSMIC STRINGS}\label{app:GWs}

\onecolumngrid
%%%%%%%%%% Merge with supplemental materials %%%%%%%%%%
\setcounter{equation}{0}
\setcounter{figure}{0}
\setcounter{table}{0}
\setcounter{section}{0}
\setcounter{page}{1}
\makeatletter
\renewcommand{\theequation}{S\arabic{equation}}
\renewcommand{\thefigure}{S\arabic{figure}}

\section{Proton decays including SUSY contributions}

We extend our discussion on proton decays including SUSY contributions. In SUSY GUTs, it is useful to impose an R-partiy to forbid unnecessary BNV renormalisable operators of SM superfields.  
These terms are stringently constrained by experimental bounds on the proton lifetime and are not directly related to the GUT scale
(see Ref.~\cite{Dreiner:1997uz} for a review). 

We have checked that imposing SUSY with R-parity does not change the main picture of breaking chains and phenomenological consequences listed in Fig.~\ref{fig:breaking_1}. The only exception is the prediction of a different topological defects in $SO(10) \to SU(5)$ in type (d) breaking chain \cite{Jeannerot:2003qv}, as indicated in the figure. Therefore, for types (a), (b) and (c), the same scale orderings between $\Lambda_{\rm pd}$ and $\Lambda_{\rm cs}$ can be obtained in both SUSY and non-SUSY versions. 

Most SUSY GUTs provide an additional source of proton decay, which may enhance the decaying channel $p\to K\bar{\nu}$. 
This new source can be described by dimension-five operators constructed from two SM fermions and two superpartners which are generated via colour-triplet Higgs mediation:
\begin{eqnarray} \label{eq:D5}
\frac{c_1}{M_T}(\tilde{Q} \tilde{Q})(\overline{L^c} Q) + \frac{c_2}{M_T} (\tilde{u}_R \tilde{d}_R)(\overline{e^c_R} u_R)\,,
\end{eqnarray}
where $c_1$ and $c_2$ are model-dependent coefficients 
and $M_T$ is the heavy colour-triplet Higgs mass which is correlated with $\Lambda_{\rm pd}$.
These operators are dressed via gluinos, charginos and neutralinos which give rise to dimension-six operators \cite{Ellis:1981tv,Sakai:1981pk,Dimopoulos:1981dw}.
The decay width with respect to these operators is suppressed not only by $M_T^2$ but also by Yukawa couplings, loop factors and the unknown SUSY-breaking scale. Therefore, the connection between the GUT scale and lifetime measured in the $K^+\bar{\nu}$ channel is weaker than that measured in the $\pi^0 e^+$ channel. 
Depending on the model, the contribution to proton decay from such SUSY GUT
operators, particularly in the $K^+ \bar{\nu}$ channel,
can be enhanced \cite{Raby:2004br,Nath:2006ut} as compared with the non-SUSY contribution. 

The experimental constraint to the $K^+\bar{\nu}$ channel has been improved to $\tau_{K^+ \bar{\nu}} > 6.6 \times 10^{33}$ years  in Super-Kamiokande at $90\%$ C.L. \cite{Miura:2016krn}. 
In the future, DUNE \cite{Acciarri:2015uup} and JUNO \cite{An:2015jdp} which will set limits at $\tau_{K^+ \bar{\nu}}  \gtrsim 5.0 \times 10^{34}$  and $3.0 \times 10^{34}$ years, respectively. 
The complementarity of these nucleon decay searches in the upcoming large-scale experiments  will provide us with an unprecedented opportunity to probe the ultra-high energy GUT scale (see {\it e.g.}, Ref.~\cite{Heeck:2019kgr}).

We note that the minimal SUSY $SU(5)$  model exhibits a tension for $M_T$ between its gauge unification prediction and constraints by Super-Kamiokande \cite{Goto:1999iz,Murayama:2001ur}, which impacts on model constructions of types (a) and (d) breaking chains. For realistic models overcoming this inconsistency, see, {\it e.g.}, Refs. \cite{Raby:2004br,Nath:2006ut}, where the dimension-five operators of \equaref{eq:D5} with specified coefficients still applies \cite{Ellis:1981tv,Nath:1985ub,Nath:1988tx,Hisano:1992jj}. 
The dimension-five operator is suppressed in flipped $SU(5)$, referring to an intermediate symmetry in type (b), due to the missing-partner mechanism  \cite{Antoniadis:1987dx}. Therefore, even in the SUSY case, type (b) may not predict an observable signature for the $K^+\bar{\nu}$ channel as mentioned in Fig.~\ref{fig:observations}.

\section{Numerical methods to calculate GWs from cosmic strings}

We list numerical methods to calculate the SGWB released from cosmic strings. Those from the general undiluted cosmic strings and from inflation-diluted ones will be considered separately. 
  
{\bf 1. GWs via undiluted cosmic strings}. 
We follow \cite{Cui:2018rwi} to estimate the emission of stochastic gravitational wave background (SGWB) from cosmic string scaling. We assume a standard cosmology and that the majority of the energy loss of the cosmic string is dominated by gravitational radiation rather than particle production, although new physics, which triggers an early period of matter domination, can affect the SGWB \cite{Cui:2017ufi,Cui:2018rwi,Gouttenoire:2019rtn,Gouttenoire:2019kij}. Considering ideal Nambu-Goto strings, the dominant radiation emission is in the form of GWs. 
Note that for cosmic strings generated from gauge symmetry breaking, energy released from the string decay may be transferred not only to gravitational radiation but also into excitations of their elementary constituents. As pointed in \cite{Auclair:2019wcv}, in the absence of long-range interactions, excitations in the vacuum are massive (which is true in GUTs) and hence the expectation is that this radiation will be suppressed for long wavelength modes of the strings. Recent simulations of the Abelian Higgs model show that the particle production is only important for extremely small loops, and therefore the gravitational wave production is dominant for most situations \cite{Matsunami:2019fss}. 
We assume there is no qualitative change for strings from gauge symmetry breaking which is also the assumption adopted in  \cite{Dror:2019syi}. However, large-scale field theory simulations of the whole network of strings show discrepancies with this statement. Due to this unsolved discrepancy, there may be large uncertainties associated with  the constraints on the cosmic string scale \cite{Hindmarsh:2017qff}. We anticipate this issue will be clarified in the coming years when  the next-generation neutrino and GW experiments take place.

For Nambu-Goto strings, the large loops give the dominant contribution to the GW signal and therefore, we focus on them. The initial large loops have typical length $l_i = \alpha t_i$ with $\alpha \simeq 0.1$ numerically obtained  \cite{Blanco-Pillado:2013qja,Blanco-Pillado:2017oxo} and $t_i$ the initial time of string formation. 
The length of loops decreases as they release energy to the cosmological background, 
\begin{eqnarray} \label{eq:length}
l(t) = l_i - \Gamma G \mu (t-t_i)
 \,.
\end{eqnarray}
Frequencies of GW released from the loops are given by $2k/l_i$ where $k=1,2,\cdots$.

We denote the $\Lambda_{\rm cs}$ as the scale of symmetry breaking leading to GW. 
The tension of the string (energy per unit length) $\mu$ is typical taken to be $\Lambda_{\rm cs}^2$. After strings form, loops are found to emit energy in the form of gravitational radiation at a constant rate
\begin{eqnarray}
\frac{dE}{dt} = - \Gamma G \mu^2 \,,
\end{eqnarray}
where numerically $\Gamma$ is found to be $\Gamma \approx 50$ \cite{Burden:1985md,Vilenkin:2000jqa,Blanco-Pillado:2017oxo}. 

Assuming the fraction of the energy transfer in the form of large loops is $\mathcal{F}_\alpha \simeq 0.1$, the relic GW density parameter is given by
\begin{eqnarray}
\Omega_{\rm GW} (f) = \frac{1}{\rho_c} \frac{d\rho_{\rm GW}}{d \log f}.
\end{eqnarray}
This can be written as a summation of mode $k$
\begin{eqnarray}
\Omega_{\rm GW}(f) = \sum_k \Omega_{\rm GW}^{(k)}(f),
\end{eqnarray}
with 
\begin{eqnarray}
\Omega_{\rm GW}^{(k)} (f) &= &\frac{1}{\rho_c} \frac{2k}{f} \frac{\mathcal{F}_\alpha \Gamma^{(k)} G \mu^2}{\alpha (\alpha +  \Gamma G \mu)} \nonumber\\
\int^{t_0}_{t_F} &dt& \frac{C_{\rm eff} (t_i^{(k)})}{t_i^{(k)4}} \frac{a^2(t) a^3(t_i^{(k)})}{ a^5(t_0)} \theta(t_i^{(k)} - t_F),
\end{eqnarray}
where $\rho_c$ is the critical energy density of the Universe given by
\begin{eqnarray}
\Gamma^{(k)} &=& \frac{1}{3.6}\Gamma k^{-4/3}, \nonumber\\
t_i^{(k)} &=& \frac{1}{\alpha + \Gamma G \mu} \left( \frac{2k}{f} \frac{a(t)}{a(t_0)} + \Gamma G \mu t \right), 
\end{eqnarray}
 $C_{\rm eff}$ is numerically obtained as 
$C_{\rm eff} = 5.7, 0.5$ \cite{BlancoPillado:2011dq,Blanco-Pillado:2017oxo,Blanco-Pillado:2013qja} for radiation and matter domination, respectively, and $t_F$ is the time of string network formation. 

In our numerical calculation of GW spectrum, we have fixed the numerical values of $\alpha$ and $\mathcal{F}_\alpha$ at values suggested by simulation in the literature. This treatment ignores uncertainties from the simulation which introduce an additional uncertainty for the prediction of GW spectrum. We anticipate these uncertainties will be under better control in the future simulation given the timeline of data-taking for the next-generation neutrino experiments and GW observatories is in the next decade.

{\bf 2. GWs of inflation-diluted cosmic strings}.
Our assumptions follow those outlined in \cite{Cui:2019kkd} where it is assumed during inflation 
the Hubble expansion rate  is constant with $H = H_{\rm I} \equiv \sqrt{V_{\rm I}/3 M_{\rm Pl}^2}$ with $V_{\rm I} = \Lambda_{\rm inf}^4$ and $M_{\rm Pl}$ the reduced Planck mass.  

Cusps on string loops lead to bursts of GWs, which can potentially be resolved as individual events \cite{Damour:2000wa,Damour:2001bk,Damour:2004kw,Siemens:2006vk}. Kinks also leads to bursts of GWs but subdominant \cite{Blanco-Pillado:2017oxo,Ringeval:2017eww}, which will not be considered here. 
Assuming the correlation length of strings as $L$, together with the speed of string $\bar{v}$,  satisfy
\begin{eqnarray}
\frac{dL}{dt} &=& (1+ \bar{v}^2) HL + \frac{\tilde{c} \bar{v}}{2} \,, \nonumber\\
\frac{d\bar{v}}{dt} &=& (1- \bar{v}^2) \left[ \frac{k(\bar{v})}{L} - 2 H \bar{v} \right] \,, 
\end{eqnarray}
where $\tilde{c}\approx 0.23$ parametrises the  loop chopping efficiency \cite{Martins:2000cs} and 
\begin{eqnarray}
k(\bar{v}) = \frac{2\sqrt{2}}{\pi} (1-\bar{v}^2) (1+2\sqrt{2}\bar{v}^3) 
\left( \frac{1-8\bar{v}^6}{1+8\bar{v}^6} \right) \,.
\end{eqnarray}
During inflation, the solution is simplified to 
\begin{eqnarray}
L(t) &=& L_F e^{H_{\rm I} ( t- t_F)} \,, \nonumber\\
\bar{v}(t) &=& \frac{2\sqrt{2}}{\pi} \frac{1}{H_{\rm I} L(t)} \,,
\end{eqnarray} 
where $L_F = L(t_F)$ is the initial condition, and $t_F$ is the network formation time (assuming it happens after the beginning of inflation). 
Inflation results in the string out of horizon $HL \gg 1$. After inflation ends, $HL$ reduces and eventually evolves back to the horizon. We denote $\tilde{z}$ as the redshift when strings returns to the horizon, $H(\tilde{z})L(\tilde{z}) = 1$. 

The rate of bursts per volume per length $d^2 R / dV dl$ observed today can be  transferred to the rate of per redshift per waveform as
\begin{equation}
\frac{d^2 R}{dz \, dh} (h,z,f) = \frac{2^{3(q-1)}\pi G \mu N_q}{2-q} \frac{r(z)}{(1+z)^5 H(z)} \frac{n(l,z)}{h^2 f^2} \,,
\label{eq:d2Rdzdh}
\end{equation}
where $h$ is the waveform, $r(z) = \int_0^z dz'/H(z')$ is the proper distance to the source and $q=4/3$ for cusps. During the inflationary era, $r(z)$ is simplified to $r(z) = r_{\rm R} + (z-z_{\rm R})/ H_{\rm I}$, where $r_{\rm R} = r (z_{\rm R})$ represents the reheating period in the end of inflation. 

$n(l,t)$ is the differential number density of long loops per unit length given by
\begin{equation}
n(l,t) = \frac{F_\alpha}{\sqrt{2}} \frac{(z(t)+1)^3/(z(t_i)+1)^3}{\alpha dL/dt|_{t=t_i} + \Gamma G \mu} \frac{\tilde{c} \bar{v}(t_i)}{\alpha L^4(t_i)} \,,
\end{equation}
where $l$ is the length of string given in \equaref{eq:length} with initial length replaced by $l_i = \alpha L(t)$. $l$ is correlated with the waveform of loops $h$. Given the redshift $z$, the frequency $f$ and $h$, $l$ is determined to be 
\begin{eqnarray}
l(h,z,f) = \left( f^q (1+z)^{q-1} h \frac{r(z)}{G \mu}\right)^{1/(2 - q)}
\end{eqnarray}
for cusps. 

From this correlation, one can determine $t_i$ for given $t$, $h$ and $f$. 
In order to ensure a solution for $t_i$, $l(h,z,f)< \alpha L(z)$ is satisfied, it is equivalent to to setting an upper bound value of $h$.
 Furthermore, the above formulas are valid only for small angle radiation, {\it i.e.}, $\theta_m = 1/(fl(1+z))^{1/3}<1$, which provides a lower bound value of $h$. 
In summary, $h$ is restricted in the interval $(h_{\rm min},~ h_{\rm max})$ with
\begin{eqnarray}
h_{\rm min} &=& \frac{1}{(1+z)f^2} \frac{G \mu}{r(z)} \,,\\
h_{\rm max} &= &\frac{[\alpha L(z)]^{2-q}}{(1+z)^{q-1} f^q} \frac{G \mu}{r(z)} \,.
\end{eqnarray}
Bursts contribute to the SGWB as
\begin{equation}
\Omega_{\rm GW}^{\rm diluted} (f) = \frac{1}{\rho_c} \frac{\pi}{2} 
f^3 
\int^{\infty}_{z_*} dz 
 \int^{h_{\rm max}}_{h_{\rm min}} dh \, h^2 \frac{d^2 R}{dz dh} (h,z,f) \,,
\end{equation}
where $z_*$ enforces the rate condition and solved via
\begin{eqnarray}
f = \int^{z_*}_0 dz 
 \int^{h_{\rm max}}_{h_{\rm min}} dh \frac{d^2 R}{dz \, dh} (h,z,f) \,.
\end{eqnarray}

\end{document}